\documentclass[graybox]{svmult}

\usepackage{type1cm}        
\usepackage{makeidx}         
\usepackage{graphicx}        
\usepackage{multicol}        
\usepackage[bottom]{footmisc}

\usepackage{braket}
\usepackage{txfonts}
\usepackage[numbers,sort&compress]{natbib}
\usepackage{hyperref}

\makeindex             

\definecolor{orange}{RGB}{20, 150, 60}

\newcommand{\eqref}[1]{(\ref{#1})}%
\newcommand{\figref}[1]{Fig.~\ref{#1}} %
\newcommand{\secref}[1]{Sec.~\ref{#1}}%
\newcommand{\dvec}[2]{\mathbf{#1}_{#2}}%
\newcommand{\uvec}[2]{\mathbf{#1}^{#2}}%
\newcommand{\dcom}[2]{{#1}_{#2}}%
\newcommand{\ucom}[2]{{#1}^{#2}}%
\newcommand{\udcom}[3]{{#1}^{#2}_{\cdot #3}}%
\newcommand{\ducom}[3]{{#1}_{#2}^{\cdot #3}}%
\newcommand{\ucombasis}[3]{\bigl[ \ucom{#1}{#2} \bigr]_\mathcal{#3}}
\newcommand{\dcombasis}[3]{\bigl[ \dcom{#1}{#2} \bigr]_\mathcal{#3}}
\newcommand{\udcombasis}[4]{\bigl[ \udcom{#1}{#2}{#3} \bigr]_\mathcal{#4}}
\newcommand{\ducombasis}[4]{\bigl[ \ducom{#1}{#2}{#3} \bigr]_\mathcal{#4}}
\newcommand{\phyd}[1]{\dim \bigl[{#1}\bigr]}

\newcommand{\covd}[2]{\nabla_{#2}{#1}}
\newcommand{\chris}[2]{\Gamma^{#1}_{#2}}
\newcommand{\pcom}[2]{{#1}({#2})}

\begin{document}
\title*{Representation of continuum equations in physical components for arbitrary curved surfaces}
\titlerunning{Continuum equations in physical components}
\author{Sujit Kumar Nath}
\institute{Sujit Kumar Nath$^{1, 2, 3}$ \at
$^1$ Aix Marseille Univ, Universit\'e de Toulon, CNRS, CPT (UMR 7332), Turing
Center for Living Systems, Marseille, France\\
$^2$ Division of Cardiovascular Sciences, Faculty of Biology, Medicine and Health,  The University of Manchester, M13 9PT, Manchester, United Kingdom \\
$^3$ School of Computing and Faculty of Biological
Sciences, University of Leeds, LS2 9JT, Leeds, United Kingdom \\
\email{sujit-kumar.nath@univ-amu.fr}}
%
%
\maketitle


\abstract{Continuum equations are ubiquitous in physical modelling of elastic, viscous, and viscoelastic systems. The equations of continuum mechanics take nontrivial forms on curved surfaces. Although the curved surface formulation of the continuum equations are derived in many excellent references available in the literature, they are not readily usable for solving physical problems due to the covariant, contravariant or mixed nature of the stress and strain tensors in the equations. We present the continuum equations in terms of physical components in a general differentiable manifold. This general formulation of the continuum equations can be used readily for modelling physical problems on arbitrary curved surfaces. We demonstrate this with the help of some examples.}

\section{Introduction}
Equations of continuum mechanics are essential tools for modelling a broad range of physical systems, starting from micro/nano-scale biological systems of cells \cite{biofilm,gaffney}, to macro-scale ecological systems \cite{tonerTu,tonerTuRama} to large scale (parsec) astrophysical systems~\cite{nath2015origin,nath2016hydro,nath2013magnetohydro,nath2014crosscorr}. However, the physical systems under consideration in various contexts are not always planar and not suitable for representing in a Cartesian coordinate system. Even in some cases, such as fluid flow on curved surfaces, it is necessary to represent the continuum equations in a way that takes care of the geometry of the curved surface. There is a plethora of excellent references in the literature which derives the continuum equations in
curved geometry~\cite{sokolnikoffTensor,lebedevTensor,audoly,altman1993}. However, most of these literatures leave the final
equations with stress, strain, displacement, velocity etc. represented in covariant, contravariant or mixed
tensor components. As a result, these equations are not readily usable for physical modelling
purpose as the dimensions of covariant, contravariant or mixed components of stress, strain,
velocity, displacement etc. do not always match with their corresponding physical
dimensions.

In this article we present continuum mechanical equations in physical components for arbitrary
curved surfaces by consolidating previous seminal works by Sokolnikoff, Truesdell, McConnell, Altman \& Oliveira etc.
\cite{sokolnikoffTensor,altman1993,altman1995,truesdell1953,mac,frederick}.
There are extensive amount of references on physical coordinates and continuum equations
on arbitrary curved surfaces, separately. However, references for continuum equations in physical components of arbitrary curved surfaces are very rare. As a result, presenting the
correct form of continuum equations for solving physical problems on curved surfaces
is very important and essential. In this article we present the continuum mechanical equations in terms of the physical components of vectors and tensors. The
main contribution of this article is to summarize the general formulation of continuum
equations which are immediately applicable in the modelling of physical systems in arbitrary curved geometry without any dimensional ambiguity.
\section{The concept of physical components of vectors and tensors}
\label{secPhysicalComponents}
Intuitively, the physical components of any scalar, vector, or tensor
are the components whose dimensions are same as their dimensions while measured during
any physical/experimental measurement. The continuum mechanical equations
are generally represented in covariant, contravariant or mixed tensor notations when
the equations are derived for arbitrary curved manifolds. Therefore, these equations are
not readily usable for physical modelling as the dimensions of the covariant, contravariant,
or mixed tensors do not match with their physically measured counterparts. Hence, it is
important to represent these equations in their correct physical components which can be
used for physical modelling purpose, being immune to any dimensional mismatch.
\subsection{Physical components of vectors}
\label{subsecPhyCompVec}
Let $\mathbf{v} = \ucom{v}{i} \dvec{g}{i}= \dcom{v}{i}\uvec{g}{i}$ be the representation of a
vector $\mathbf{v}$ in its covariant and contravariant bases $\{\dvec{g}{i}\}$ and $\{\uvec{g}{i}\}$
with components $\ucom{v}{i}$ and $\dcom{v}{i}$ respectively.
To demonstrate the problem at hand with an example, let us consider the dynamics of a particle in plane polar coordinate. The parametrization of the $2$D-plane in polar coordinate is given by
$\mathbf{P}=(r \cos\theta, r \sin\theta)$, where $r$ is the radial distance of the point $\mathbf{P}$
from the origin
and $\theta$ is the angular coordinate. In this coordinate system the contravariant
and covariant components of the acceleration $\mathbf{a}=\ucom{a}{i}\dvec{g}{i}=\dcom{a}{i}\uvec{g}{i},
i\in \{r, \theta\}$ can be written as
\begin{eqnarray}
\label{eqPolarAccConComp}
\ucom{a}{r} = \frac{d^2r}{dt^2}-r\left(\frac{d\theta}{dt}\right)^2, \,\,\,
\ucom{a}{\theta} = \frac{d^2\theta}{dt^2}+\frac{2}{r}\frac{dr}{dt}\frac{d\theta}{dt}, \\
\label{eqPolarAccCovComp}
\dcom{a}{r} = \frac{d^2r}{dt^2}-r\left(\frac{d\theta}{dt}\right)^2, \,\,\,
\dcom{a}{\theta} = r^2\frac{d^2\theta}{dt^2}+2r\frac{dr}{dt}\frac{d\theta}{dt}.
\end{eqnarray}
We see in \eqref{eqPolarAccConComp} and \eqref{eqPolarAccCovComp} that the dimensions of
$\ucom{a}{r},\ucom{a}{\theta}, \dcom{a}{r}$ and $\dcom{a}{\theta}$ are LT$^{-2}$, T$^{-2}$, LT$^{-2}$ and
L$^2$T$^{-2}$, respectively. Clearly, not all the dimensions are same as the dimension of acceleration.
Even dim$\bigl[\ucom{a}{\theta}\bigr]$ and dim$\bigl[\dcom{a}{\theta}\bigr]$ also differ from each other.

\begin{figure}
\includegraphics[scale=0.33]{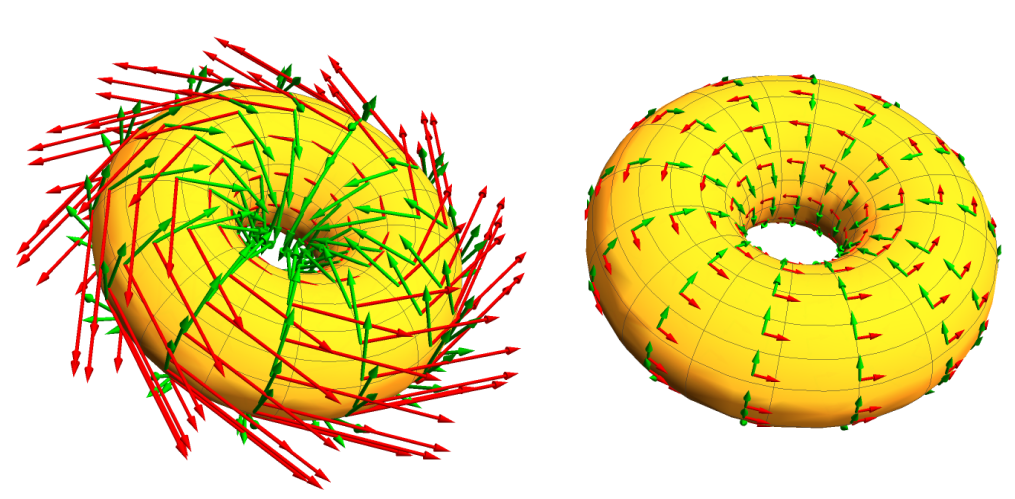}
\caption{(Color online) Basis vectors on the surface of a torus parametrized by
$\mathbf{P}=\left[(a + b\cos\theta) \cos\phi, (a + b\cos\theta) \sin\phi, b\sin\theta \right]$,
where $a$ and $b$ are the radii of the major and minor circles respectively, and $\theta, \phi$ are
the angular parameters.
Left panel shows the covariant basis vectors $\dvec{g}{\phi}$ (red arrows) and $\dvec{g}{\theta}$
(green arrows) without normalization. Right panel shows the normalized basis vectors
$\dvec{e}{\phi}=\dvec{g}{\phi}/\sqrt{\dcom{g}{\phi\phi}}$ (red)
and $\dvec{e}{\theta}=\dvec{g}{\theta}/\sqrt{\dcom{g}{\theta\theta}}$ (green) after normalization.}
\label{figTorus}
\end{figure}

To circumvent this dimensional mismatch, it is necessary to represent the vectors and tensors
in their physical components while modelling physical systems. In this purpose, it is useful
to consider the unit vectors
\begin{equation}
\label{eq1stUnitVec}
\dvec{e}{i}=\frac{1}{\sqrt{\dcom{g}{ii}}}~\dvec{g}{i}, \,\,\, \uvec{f}{i}=\frac{1}{\sqrt{\ucom{g}{ii}}}~\uvec{g}{i},\\
\end{equation}
and their duals
\begin{equation}
\label{eq2ndUnitVec}
\uvec{e}{i}=\sqrt{\dcom{g}{ii}}~\uvec{g}{i}, \,\,\, \dvec{f}{i}=\sqrt{\ucom{g}{ii}}~\dvec{g}{i}
\end{equation}
to construct the bases, instead of using the covariant or contravariant bases directly.
The covariant basis vectors and their normalized counterparts are shown for an example case of a $2-$dimensional toroidal surface in \figref{figTorus}.
Note that the dual basis vectors in \eqref{eq2ndUnitVec} are not
necessarily unit vectors in general curvilinear coordinates (although they are unit vectors in orthogonal curvilinear coordinates).
Any arbitrary vector $\mathbf{v}$ can now be written as
\begin{equation}
\label{eqPhyComp}
 \mathbf{v} = \ucombasis{v}{i}{E}~\dvec{e}{i}
 = \dcombasis{v}{i}{E^*}~\uvec{e}{i}
 = \ucombasis{v}{i}{F}~\dvec{f}{i}
  = \dcombasis{v}{i}{F^*}~\uvec{f}{i},
\end{equation}
where the notations $\mathcal{E}, \mathcal{E^*}, \mathcal{F}, \mathcal{F^*}$ denote the bases
$\{\dvec{e}{i}\}, \{\uvec{e}{i}\}, \{\dvec{f}{i}\}$ and $\{\uvec{f}{i}\}$ respectively, and 
the components $\bigl[\ucom{v}{i}\bigr]_\alpha,\,\alpha\in\{\mathcal{E}, \mathcal{F}\}$,
and $\bigl[\dcom{v}{i}\bigr]_\alpha,\,\alpha\in\{\mathcal{E^*}, \mathcal{F^*}\}$
in \eqref{eqPhyComp} are called the physical components
of $\mathbf{v}$ in representative bases. Therefore, we obtain the relation between the 
physical components and the covariant or contravariant components as
\begin{eqnarray}
\label{eqPhyCompVec}
\ucombasis{v}{i}{E} = \sqrt{\dcom{g}{ii}}~\ucom{v}{i}, \,\,
%
\ucombasis{v}{i}{F} = \frac{1}{\sqrt{\ucom{g}{ii}}}~\ucom{v}{i}, \,\,
%
\dcombasis{v}{i}{E^*} = \frac{1}{\sqrt{\dcom{g}{ii}}}~\dcom{v}{i}, \,\,
%
\dcombasis{v}{i}{F^*} = \sqrt{\ucom{g}{ii}}~\dcom{v}{i}.
\end{eqnarray}
Note that in the above formulae in \eqref{eqPhyCompVec} the right hand sides (RHS) are not
summed over the repeated index $i$. Since for orthogonal curvilinear coordinates
$\dcom{g}{ii}=1/\ucom{g}{ii}$, it is clear from \eqref{eqPhyCompVec} that
$\ucombasis{v}{i}{E}=\ucombasis{v}{i}{F}$ and
$\dcombasis{v}{i}{E^*}=\dcombasis{v}{i}{F^*}$ in orthogonal
curvilinear coordinates. However,
for oblique coordinates these equalities do not necessarily hold. Therefore, it is necessary
to choose the correct physical components while modelling physical systems using continuum
equations. In plane polar coordinate we have $\dcom{g}{rr}=\ucom{g}{rr}=1$,
$\dcom{g}{\theta\theta}=r^2$ and $=\ucom{g}{\theta\theta}=1/r^2$. Therefore, considering our
example in \eqref{eqPolarAccConComp} and \eqref{eqPolarAccCovComp} we obtain
\begin{eqnarray}
\ucombasis{a}{r}{E} = \sqrt{\dcom{g}{rr}}~\ucom{a}{r} = \ucom{a}{r}, \, \,
\ucombasis{a}{\theta}{E} = \sqrt{\dcom{g}{\theta\theta}}~\ucom{a}{\theta} = r~\ucom{a}{\theta}, \\
\dcombasis{a}{r}{E^*} = \frac{1}{\sqrt{\dcom{g}{rr}}}~\dcom{a}{r} = \dcom{a}{r}, \, \,
\dcombasis{a}{\theta}{E^*} = \frac{1}{\sqrt{\dcom{g}{\theta\theta}}}~\dcom{a}{\theta}
=\frac{\dcom{a}{\theta}}{r},
\end{eqnarray}
It is now evident that all the physical components $\ucombasis{a}{r}{E}, \ucombasis{a}{\theta}{E},
\dcombasis{a}{r}{E^*},$ and $\dcombasis{a}{\theta}{E^*}$ have the dimension of acceleration.
Also the same is true for the physical components
$\ucombasis{a}{r}{F}, \ucombasis{a}{\theta}{F}, \dcombasis{a}{r}{F^*},$ and
$\dcombasis{a}{\theta}{F^*}$. In the next section we show that for a general tensor of any 
arbitrary order all the physical components carry the same physical dimension. 
\subsection{Physical components of tensors}
We shall discuss here the physical components of second order tensors only. The case of tensors
of any arbitrary order can easily be generalised from this discussion. Let
$\mathbb{T} = \ucom{\tau}{ij}\dvec{g}{i}\otimes\dvec{g}{j}
= \dcom{\tau}{ij}\uvec{g}{i}\otimes\uvec{g}{j}
= \udcom{\tau}{i}{j}\dvec{g}{i}\otimes\uvec{g}{j}
= \ducom{\tau}{i}{j}\uvec{g}{i}\otimes\dvec{g}{j}$ be a second order tensor, represented
in the covariant, contravariant, and mixed bases. We can represent the physical components
in the same way we did for vectors by representing the tensor components in unit bases.
Therefore, the $16$ physical components of the tensor $\mathbb{T}$ are given by 
\begin{eqnarray}
\label{eqPhyCompTensor}
\nonumber
\ucombasis{\tau}{ij}{E\otimes E} = \sqrt{\dcom{g}{ii} \dcom{g}{jj}}~\ucom{\tau}{ij},\,\,
\ucombasis{\tau}{ij}{E\otimes F} = \sqrt{\frac{\dcom{g}{ii}}{\ucom{g}{jj}}}~\ucom{\tau}{ij},\\
\nonumber
\ucombasis{\tau}{ij}{F\otimes E} = \sqrt{\frac{\dcom{g}{jj}}{\ucom{g}{ii}}}~\ucom{\tau}{ij}, \,\,
\ucombasis{\tau}{ij}{F\otimes F} = \frac{1}{\sqrt{\ucom{g}{ii} \ucom{g}{jj}}}~\ucom{\tau}{ij},\\
\nonumber
\udcombasis{\tau}{i}{j}{E\otimes E^*} = \sqrt{\frac{\dcom{g}{ii}}{\dcom{g}{jj}}}~\udcom{\tau}{i}{j},\,\,
\udcombasis{\tau}{i}{j}{E\otimes F^*} = \sqrt{\dcom{g}{ii}\ucom{g}{jj}}~\udcom{\tau}{i}{j},\\
\nonumber
\udcombasis{\tau}{i}{j}{F\otimes E^*} = \frac{1}{\sqrt{\ucom{g}{ii} \dcom{g}{jj}}}~\udcom{\tau}{i}{j}, \,\,
\udcombasis{\tau}{i}{j}{F\otimes F^*} = \sqrt{\frac{\ucom{g}{jj}}{\ucom{g}{ii}}}~\udcom{\tau}{i}{j},\\
\nonumber
\ducombasis{\tau}{i}{j}{E^*\otimes E} = \sqrt{\frac{\dcom{g}{jj}}{\dcom{g}{ii}}}~\ducom{\tau}{i}{j},\,\,
\ducombasis{\tau}{i}{j}{E^*\otimes F} = \frac{1}{\sqrt{\dcom{g}{ii} \ucom{g}{jj}}}~\ducom{\tau}{i}{j},\\
\nonumber
\ducombasis{\tau}{i}{j}{F^*\otimes E} = \sqrt{\ucom{g}{ii}\dcom{g}{jj}}~\ducom{\tau}{i}{j},\,\,
\ducombasis{\tau}{i}{j}{F^*\otimes F} = \sqrt{\frac{\ucom{g}{ii}}{\ucom{g}{jj}}}~\ducom{\tau}{i}{j},\\
\nonumber
\dcombasis{\tau}{ij}{E^*\otimes E^*} = \frac{1}{\sqrt{\dcom{g}{ii} \dcom{g}{jj}}}~\dcom{\tau}{ij},\,\,
\dcombasis{\tau}{ij}{E^*\otimes F^*} = \sqrt{\frac{\ucom{g}{jj}}{\dcom{g}{ii}}}~\dcom{\tau}{ij},\\
%
\dcombasis{\tau}{ij}{F^*\otimes E^*} = \sqrt{\frac{\ucom{g}{ii}}{\dcom{g}{jj}}}~\dcom{\tau}{ij},\,\,
\dcombasis{\tau}{ij}{F^*\otimes F^*} = \sqrt{\ucom{g}{ii} \ucom{g}{jj}}~\dcom{\tau}{ij}.
\end{eqnarray}
It can be shown that all the physical components of a tensor have the same dimension 
which is same as their dimension in Cartesian coordinate~\cite{altman1995}.
However, the method opted in~\cite{altman1995} is not always applicable for tensor 
fields on surfaces of codimension one, as there may not always exist a suitable Cartesian
coordinate for representing a tensor field on such a surface. Here we present an alternative
way to prove that all the physical components of a tensor have the same dimension with 
the help of intrinsic geometry of the surface. We again demonstrate this with the 
help of second order tensors, which can be easily generalised for tensors of arbitrary 
order. 

We first derive a few results relating the physical dimensions of the covariant and 
contravariant metric tensor and vector components. Considering the squared arc length
\begin{equation}
\label{eqarclen}
 ds^2 = \dcom{g}{ij}\,\ucom{dx}{i}\ucom{dx}{j} = \ucom{g}{ij}\,\dcom{dx}{i}\dcom{dx}{j} 
 = \ucom{dx}{i}\dcom{dx}{i}
\end{equation}
we obtain 
\begin{eqnarray}
\label{eqphyddgij}
\phyd{\dcom{g}{ij}} &=& \frac{L^2}{\phyd{\ucom{dx}{i}} \phyd{\ucom{dx}{j}}}, \\ 
\label{eqphydugij}
\phyd{\ucom{g}{ij}} &=& \frac{L^2}{\phyd{\dcom{dx}{i}} \phyd{\dcom{dx}{j}}}, \\
\label{eqphydxi}
\phyd{\ucom{dx}{i}} &=& \frac{L^2}{\phyd{\dcom{dx}{i}}}, 
\end{eqnarray}
where $L$ denotes the dimension of arc length $ds$ in \eqref{eqarclen}. Substituting the $\phyd{\ucom{dx}{i}}$, 
$\phyd{\ucom{dx}{j}}$ from \eqref{eqphydxi} to \eqref{eqphyddgij} and comparing with 
\eqref{eqphydugij} we obtain 
\begin{equation}
\label{eqphydgijinverse}
 \phyd{\dcom{g}{ij}} = \frac{1}{\phyd{\ucom{g}{ij}}}.
\end{equation}
Again, from \eqref{eqarclen} we see that
\begin{equation}
\label{eqPhydDgiivsUxi}
 \phyd{\sqrt{\dcom{g}{ii}}} = \frac{L}{\phyd{\ucom{dx}{i}}}, 
\end{equation}
and 
\begin{equation}
\label{eqPhydUgiivsDxi}
 \phyd{\sqrt{\ucom{g}{ii}}} = \frac{L}{\phyd{\dcom{dx}{i}}}.
\end{equation}
Therefore, \eqref{eqphyddgij} and \eqref{eqphydugij} gives 
\begin{equation}
\label{eqgijvsgiigjj1}
 \phyd{\dcom{g}{ij}} = \phyd{\sqrt{\dcom{g}{ii}}} \, \phyd{\sqrt{\dcom{g}{jj}}},
\end{equation}
and  
\begin{equation}
\label{eqgijvsgiigjj2}
 \phyd{\ucom{g}{ij}} = \phyd{\sqrt{\ucom{g}{ii}}} \, \phyd{\sqrt{\ucom{g}{jj}}}.
\end{equation}
We now use the results \eqref{eqarclen}-\eqref{eqgijvsgiigjj2} for proving the
equality of the physical dimensions of all the physical components of a tensor. First we show 
that all the physical components of a tensor in a fixed basis have the same 
physical dimension. For this purpose, without loss of generality, we fix our basis to be 
$\mathcal{E}\otimes\mathcal{F^*}$. Let us now consider the scalar value 
$\udcom{\tau}{i}{j} \, \dcom{dx}{i} \ucom{dx}{j}$. For any fixed values of $i$ and $j$, such as 
$\udcom{\tau}{p}{q} \, \dcom{dx}{p} \ucom{dx}{q}$, when $i=p, j=q$ fixed, and 
$\udcom{\tau}{m}{n} \, \dcom{dx}{m} \ucom{dx}{n}$, when $i=m, j=n$ fixed, we have 
\begin{equation}
\label{eqEqualitySameBaseTerms}
 \phyd{\udcom{\tau}{p}{q} \, \dcom{dx}{p} \ucom{dx}{q}} 
 = \phyd{\udcom{\tau}{m}{n} \, \dcom{dx}{m} \ucom{dx}{n}}. 
\end{equation}
The dimensions of the both sides of \eqref{eqEqualitySameBaseTerms} match as they are the typical
constitutive terms of the scalar value of $\mathbb{T}(\mathbf{u},\mathbf{v})$, i.e., when  $\mathbb{T}$ is applied
on a pair of vectors $(\mathbf{u},\mathbf{v})$.
Substituting the values of $\phyd{\dcom{dx}{p}},\phyd{\ucom{dx}{q}},\phyd{\dcom{dx}{m}},\phyd{\ucom{dx}{n}}$ in \eqref{eqEqualitySameBaseTerms} using \eqref{eqPhydDgiivsUxi} and \eqref{eqPhydUgiivsDxi} we obtain
\begin{equation}
\label{eqEqualitySameBaseCalc1} 
 \phyd{\udcom{\tau}{p}{q}} \, \frac{L}{\phyd{\sqrt{\ucom{g}{pp}}}} \, \frac{L}{\phyd{\sqrt{\dcom{g}{qq}}}}
 = \phyd{\udcom{\tau}{m}{n}} \, \frac{L}{\phyd{\sqrt{\ucom{g}{mm}}}} \, \frac{L}{\phyd{\sqrt{\dcom{g}{nn}}}}. 
\end{equation}
Now, using \eqref{eqphydgijinverse} and \eqref{eqPhyCompTensor} we obtain from \eqref{eqEqualitySameBaseCalc1}
\begin{eqnarray}
\label{eqEqualitySameBase}
\nonumber
&&\phyd{\udcom{\tau}{p}{q}} \, \phyd{\sqrt{\dcom{g}{pp}}} \, \phyd{\sqrt{\ucom{g}{qq}}} = \phyd{\udcom{\tau}{m}{n}} \, \phyd{\sqrt{\dcom{g}{mm}}} \, \phyd{\sqrt{\ucom{g}{nn}}}, \\ 
\mathrm{or,} 
&&\phyd{\udcombasis{\tau}{p}{q}{E \otimes F^*}} = \phyd{\udcombasis{\tau}{m}{n}{E \otimes F^*}}. 
\end{eqnarray}
We now show the dimensional equality of the physical tensor components represented in different 
bases such as $\mathcal{E}\otimes\mathcal{F^*}$ and $\mathcal{E}\otimes\mathcal{E}$. From the relation 
\begin{equation}
 \phyd{\udcom{\tau}{p}{q} \, \dcom{dx}{p} \ucom{dx}{q}} 
 = \phyd{\ucom{\tau}{mn} \, \dcom{dx}{m} \dcom{dx}{n}}, 
\end{equation}
and following the similar calculation as in \eqref{eqEqualitySameBaseCalc1}  and using 
\eqref{eqphydgijinverse} and \eqref{eqPhyCompTensor} we obtain
\begin{eqnarray}
\label{eqEqualitySameBaseCalc2}
\nonumber
&&\phyd{\udcom{\tau}{p}{q}} \, \phyd{\sqrt{\dcom{g}{pp}}} \, \phyd{\sqrt{\ucom{g}{qq}}} = \phyd{\ucom{\tau}{mn}} \, \phyd{\sqrt{\dcom{g}{mm}}} \, \phyd{\sqrt{\dcom{g}{nn}}}, \\ 
\mathrm{or,} 
&&\phyd{\udcombasis{\tau}{p}{q}{E \otimes F^*}} = \phyd{\ucombasis{\tau}{mn}{E \otimes E}}. 
\end{eqnarray}
Similarly, all other physical components (exhaustively listed in \eqref{eqPhyCompTensor} for a 
second order tensor) can be shown to have the same physical dimension. 
Note that the above derivations are independent of any embedding Cartesian coordinate, unlike 
the derivations given in \cite{altman1995} (for tensors) and \cite{truesdell1953} (for vectors). 
Therefore, the derivations in this article are intrinsic in nature and can be applied to any
arbitrary curved surface, such as a $2-$dimensional surface of a torus, where the methods 
followed in \cite{altman1995} and \cite{truesdell1953} might not be applicable. We now 
demonstrate an application of the physical components of vectors and tensors in the
modelling of physical systems using the equations of continuum mechanics. 
\section{Equation of motion of an elastic body in physical components}
\label{secEOMelastic}
We have seen in \secref{secPhysicalComponents} that all the physical components
of a tensor carry the same physical dimension irrespective of their representative bases. However, the basis vectors in $\mathcal{E^*}=\{\uvec{e}{i}\}$ and
$\mathcal{F}=\{\dvec{f}{i}\}$ do not have unit norm in general, unlike the vectors in the bases
$\mathcal{E}=\{\dvec{e}{i}\}$ and $\mathcal{F^*}=\{\uvec{f}{i}\}$ which have unit norm. Therefore,
it is customary to use the bases $\mathcal{E}=\{\dvec{e}{i}\}$ and 
$\mathcal{F^*}=\{\uvec{f}{i}\}$ for defining the components of physically measurable vectors 
or tensors. 

We now propose a method to obtain the continuum equations in physical components,
directly from their general covariant or contravariant form. Here we use the bases 
$\mathcal{E}=\{\dvec{e}{i}\}$ and $\mathcal{F^*}=\{\uvec{f}{i}\}$ (with basis vectors of 
unit norm) to obtain the physical components. Our formalism here is motivated by the previous work
due to Altman \& Oliveira~\cite{altman1993}.
We know from standard elasticity theory
that the equation relating the strain and displacement (for small displacements, neglecting the nonlinear terms) is given by
\begin{equation}
\label{eqStrainVsDisp}
\dcom{\epsilon}{ij} = \frac{1}{2} \left(\frac{\partial u_i}{\partial x^j} + 
\frac{\partial u_j}{\partial x^i} \right) - \chris{\alpha}{ij} u_\alpha, 
\end{equation}
where $\dcom{\epsilon}{ij}$ is the strain component, $\dcom{u}{i}$ is the displacement component,
and $\Gamma^\alpha_{ij}$ is the Christoffel symbol of second kind. We also have the equation of motion (EOM)
in a general coordinate as  
\begin{equation}
\label{eqEOM}
\covd{\ucom{\tau}{ij}}{j} + \ucom{F}{i} = \rho \, \ucom{a}{i}, 
\end{equation}
where $\ucom{\tau}{ij}$ is the stress, $\ucom{F}{i}$ is the force per unit volume, 
$\ucom{a}{i}$ is the acceleration, and $\nabla$ denotes the covariant derivative. 
Finally, we also have the constitutive relation between stress and strain which, in case of linear
elasticity of homogeneous and isotropic media, can be written as
\begin{equation}
\label{eqConstitutiveRel}
\udcom{\tau}{i}{j} = \lambda \, \Theta \, \udcom{\delta}{i}{j} + 2\mu \udcom{\epsilon}{i}{j},
\end{equation}
where $\lambda,\mu$ are the Lam\'e's constants and $\Theta=\udcom{\epsilon}{i}{i}$ is the dilation.
However, as we have discussed in \secref{subsecPhyCompVec}, the dimensions of the 
vector or tensor quantities $\dcom{u}{i}, \ucom{a}{i}, \dcom{\epsilon}{ij}, \ucom{\tau}{ij},
\udcom{\tau}{i}{j}$
etc. are not the same as they are realised during physical/experimental measurements. Hence,
it is necessary to convert the equations \eqref{eqStrainVsDisp}- \eqref{eqConstitutiveRel} in terms
of physical components so that they can be applied to model physical systems. It can be 
easily understood that for rectangular Cartesian coordinate this conversion is not necessary 
as the covariant, contravariant or mixed components of tensors coincide with their physical 
components automatically. However, in general, even in the cases of orthogonal curvilinear
coordinates, the conversion of the equations \eqref{eqStrainVsDisp}-\eqref{eqConstitutiveRel}
in physical components is necessary to avoid any dimensional mismatch.

Denoting the physical components by $\pcom{u}{i}, \pcom{F}{i}, \pcom{\epsilon}{ij}, 
\pcom{\tau}{ij}$ etc. with respect to the bases $\mathcal{E}=\{\dvec{e}{i}\}$, 
$\mathcal{F^*}=\{\uvec{f}{i}\}$ and using the relations in \eqref{eqPhyCompVec} and 
\eqref{eqPhyCompTensor} we obtain the strain-displacement relation \eqref{eqStrainVsDisp} in 
their corresponding physical components as 
\begin{equation}
\label{eqStrainVsDispPhyComp}
\pcom{\epsilon}{ij} = 
\frac{\sqrt{\ucom{g}{ii}\ucom{g}{jj}}}{2} \left[\frac{\partial}{\partial x^j} \left(\frac{\pcom{u}{i}}{\sqrt{\ucom{g}{ii}}} \right) + \frac{\partial}{\partial x^i} \left( \frac{\pcom{u}{j}}{\sqrt{\ucom{g}{jj}}} \right) \right] 
- \sqrt{\ucom{g}{ii}\ucom{g}{jj}} \, \chris{\alpha}{ij} \frac{\pcom{u}{\alpha}}{\sqrt{\ucom{g}{\alpha\alpha}}},  
\end{equation}
where no summation is implied over the same type of indices, i.e., the metric tensor components $\ucom{g}{ii},\ucom{g}{jj}$ etc.
are not summed over $i$ or $j$. 
Next, expanding the covariant derivative in \eqref{eqEOM} we obtain 
\begin{equation}
\label{eqEOM1}
\frac{\partial\ucom{\tau}{ij}}{\partial \ucom{x}{j}} + \chris{i}{j\alpha}\ucom{\tau}{\alpha j} 
+ \chris{j}{j\alpha}\ucom{\tau}{i \alpha} + \ucom{F}{i} = \rho \, \ucom{a}{i}.  
\end{equation}
Replacing the contravariant type components of the stress tensor $\ucom{\tau}{ij}, \ucom{\tau}{i\alpha}$ etc. in terms of
physical components (using the relations in \eqref{eqPhyCompTensor}) we obtain the equation of
motion as 
\begin{equation}
\label{eqEOMPhyComp}
\frac{\partial}{\partial \ucom{x}{j}} \left(\frac{\pcom{\tau}{ij}}{\sqrt{\dcom{g}{ii}\dcom{g}{jj}}}\right) 
+ \chris{i}{j\alpha} \left(\frac{\pcom{\tau}{\alpha j}}{\sqrt{\dcom{g}{\alpha\alpha}\dcom{g}{jj}}}\right)
+ \chris{j}{j\alpha} \left(\frac{\pcom{\tau}{i\alpha}}{\sqrt{\dcom{g}{ii}\dcom{g}{\alpha\alpha}}}\right) 
+ \frac{\pcom{F}{i}}{\sqrt{\dcom{g}{ii}}} = \rho \, \frac{\pcom{a}{i}}{\sqrt{\dcom{g}{ii}}}, 
\end{equation}
where the summations are applicable only over common indices of different type (covariant or
contravariant as determined by \eqref{eqEOM1}). Similarly, we obtain the constitutive relation
between stress and strain in terms of physical components as
\begin{equation}
\label{eqConstitutiveRelPhyCom}
\pcom{\tau}{ij} = \lambda \, \left[ \sum_{i=1}^{n} \frac{\pcom{\epsilon}{ii}}{\sqrt{\dcom{g}{ii}\ucom{g}{ii}}} \right] \, \pcom{\delta}{ij}
+ 2 \, \mu \, \pcom{\epsilon}{ij},
\end{equation}
where $n$ is the dimension of the manifold. It is to be noted that for orthogonal coordinates
the constitutive relation in physical components retains the same form as its
mixed tensor form \eqref{eqConstitutiveRel}.
Equations \eqref{eqStrainVsDispPhyComp},
\eqref{eqEOMPhyComp}, and \eqref{eqConstitutiveRelPhyCom} are the general form of the
strain-displacement relation, EOM, and constitutive relation of an elastic body in
physical components in any arbitrary curvilinear coordinate, which are not only limited to orthogonal coordinates.

\begin{figure}
\includegraphics[scale=0.4]{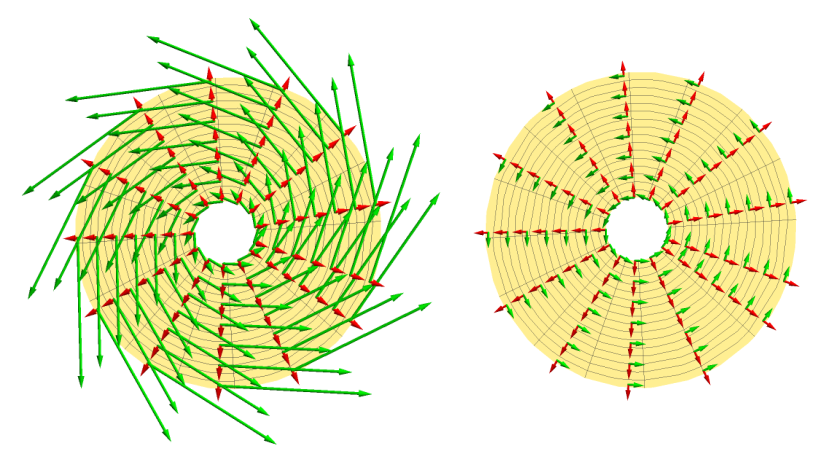}
\caption{(Color online) Basis vectors in the plane polar coordinate  parametrized by
$\mathbf{P}=\left(r\cos\theta, r\sin\theta \right)$,
where $r$ and $\theta$ are the radial and angular coordinates respectively.
Left panel shows the covariant basis vectors $\dvec{g}{r}$ (red arrows) and $\dvec{g}{\theta}$
(green arrows) without normalization. Right panel shows the normalized basis vectors
$\dvec{e}{r}=\dvec{g}{r}/\sqrt{\dcom{g}{rr}}$ (red)
and $\dvec{e}{\theta}=\dvec{g}{\theta}/\sqrt{\dcom{g}{\theta\theta}}$ (green).}
\label{figPolar}
\end{figure}

We next demonstrate, with an example of plane polar coordinate, how \eqref{eqStrainVsDispPhyComp}, \eqref{eqEOMPhyComp} and \eqref{eqConstitutiveRelPhyCom}
can be used to readily obtain the EOM of an elastic body in  physical components.
For a plane polar coordinate we have the parametrization of a $2-$dimensional plane as 
$\mathbf{P}(r,\theta)=(r\cos\theta, r\sin\theta)$, where $\mathbf{P}$ is the position vector, 
$r$ is the radial coordinate, and $\theta$ is the angular coordinate. The covariant basis vectors 
are 
\begin{equation}
\label{eqCovBasisPlanePolar}
\dvec{g}{r}=\left( \cos\theta, \sin\theta \right) \, \, \, \mathrm{and} \,\,\, 
\dvec{g}{\theta}=\left( -r\sin\theta, r\cos\theta \right). 
\end{equation}
The basis vectors $\dvec{g}{r},\dvec{g}{\theta}$ and their normalised
counterparts are shown in \figref{figPolar}.
The metric tensor $(\dcom{g}{ij})$ and its inverse $(\ucom{g}{ij})$, where $i,j\in\{r,\theta \}$, are given by
\begin{eqnarray}
\label{eqMetricPlanePolar}
\left( {\begin{array}{cc}
   \dcom{g}{rr}\, & \, \dcom{g}{r\theta} \\[2mm]
   \dcom{g}{\theta r}\, & \, \dcom{g}{\theta\theta} 
 \end{array} } \right)
 =\left( {\begin{array}{cc}
   1\, & \, 0 \\[2mm]
   0\, & \, r^2 
 \end{array} } \right)
 \,\,\, \mathrm{and} \,\,\, 
\left( {\begin{array}{cc}
   \ucom{g}{rr}\, & \, \ucom{g}{r\theta} \\[2mm]
   \ucom{g}{\theta r}\, & \, \ucom{g}{\theta\theta} 
 \end{array} } \right)
 = \left( {\begin{array}{cc}
   1\, & \, 0 \\[2mm]
   0\, & \, \frac{1}{r^2} 
 \end{array} } \right). 
\end{eqnarray}
Therefore, the dual basis vectors are given by 
\begin{equation}
\label{eqConBasisPlanePolar}
\uvec{g}{r}=\ucom{g}{rj}\dvec{g}{j}=\left( \cos\theta, \sin\theta \right) \, \, \, 
\mathrm{and} \,\,\, 
\uvec{g}{\theta}=\ucom{g}{\theta j}\dvec{g}{j}=\left( -\frac{\sin\theta}{r}, 
\frac{\cos\theta}{r} \right).
\end{equation}
The Christoffel symbols are calculated to be 
\begin{equation}
\label{eqChrisPlanePolar}
\chris{r}{\theta\theta}=-r, \chris{\theta}{r\theta}=\frac{1}{r} \,\,\,
\mathrm{and} \,\,\, 
\chris{r}{rr}=\chris{r}{r\theta}=\chris{\theta}{\theta\theta}=\chris{\theta}{rr}=0. 
\end{equation}
Therefore, using \eqref{eqStrainVsDispPhyComp}, we obtain the strain-displacement relations 
in physical components for plane polar coordinate as
\begin{eqnarray}
\label{eqStrainVsDispPhyCompPlanePolar1}
%
\pcom{\epsilon}{rr}&=&\frac{\partial\pcom{u}{r}}{\partial r}, \\ 
\label{eqStrainVsDispPhyCompPlanePolar2}
\pcom{\epsilon}{r\theta}&=&\frac{1}{2r}\left[ \frac{\partial \pcom{u}{r}}{\partial \theta} 
- \pcom{u}{\theta} + r\frac{\partial \pcom{u}{\theta}}{\partial r}\right], \\
\label{eqStrainVsDispPhyCompPlanePolar3}
\pcom{\epsilon}{\theta\theta}&=&\frac{1}{r}\left[\frac{\partial \pcom{u}{\theta}}{\partial\theta} 
- \pcom{u}{r} \right].
\end{eqnarray}
Similarly, we obtain the EOM in physical components directly from \eqref{eqEOMPhyComp} 
as 
\begin{eqnarray}
\label{eqEOMPhyCompPlanePolar1}
%
&&\frac{\partial\pcom{\tau}{rr}}{\partial r} + \frac{1}{r} 
\frac{\partial\pcom{\tau}{r\theta}}{\partial\theta} + \frac{\pcom{\tau}{rr}-\pcom{\tau}{\theta\theta}}{r} + \pcom{F}{r}
=\rho \pcom{a}{r}, \\ 
\label{eqEOMPhyCompPlanePolar2}
&&\frac{\partial\pcom{\tau}{r\theta}}{\partial r} + \frac{1}{r} 
\frac{\partial\pcom{\tau}{\theta\theta}}{\partial\theta} + \frac{2\pcom{\tau}{r\theta}}{r} 
+ \pcom{F}{\theta}
=\rho\pcom{a}{\theta}.
\end{eqnarray}
We obtain the constitutive relations in physical components from \eqref{eqConstitutiveRelPhyCom}
as
\begin{eqnarray}
\label{eqConstitutiveRelPhyComPlanePolar1}
&&\pcom{\tau}{rr} = \lambda \, \left[ \pcom{\epsilon}{rr}+\pcom{\epsilon}{\theta\theta} \right]
+ 2 \, \mu \, \pcom{\epsilon}{rr}, \\
\label{eqConstitutiveRelPhyComPlanePolar2}
&&\pcom{\tau}{r\theta} = 2 \, \mu \, \pcom{\epsilon}{r\theta}, \\
\label{eqConstitutiveRelPhyComPlanePolar3}
&&\pcom{\tau}{\theta\theta}
= \lambda \, \left[ \pcom{\epsilon}{rr}+\pcom{\epsilon}{\theta\theta} \right] + 2 \, \mu \, \pcom{\epsilon}{\theta\theta}.
\end{eqnarray}
We see from \eqref{eqStrainVsDispPhyCompPlanePolar1}-\eqref{eqConstitutiveRelPhyComPlanePolar3} that
the strain-displacement relations, EOM, and the constitutive relations in plane polar coordinate obtained from
\eqref{eqStrainVsDispPhyComp}, \eqref{eqEOMPhyComp}, and \eqref{eqConstitutiveRelPhyCom} clearly match with the results
in the literature \cite{sokolnikoffElasticity} but obtained via different procedures.
The advantage of having equations \eqref{eqStrainVsDispPhyComp}, \eqref{eqEOMPhyComp}, and
\eqref{eqConstitutiveRelPhyCom}
is that they can be applied to any curvilinear coordinate or curved surfaces readily, 
which are also not limited to the orthogonal coordinates.
The strain-displacement relationship and EOM have been derived for orthogonal 
curvilinear coordinates via different methods in seminal works by Sokolnikoff, Timoshenko etc.~\cite{sokolnikoffElasticity,timo}. However,
the general form of strain-displacement relationship, EOM, and the constitutive relation in physical components as given in \eqref{eqStrainVsDispPhyComp}, \eqref{eqEOMPhyComp}, and
\eqref{eqConstitutiveRelPhyCom} are very scarce in the previous literature.
\section{Conclusion}
\label{conclusion}
In this article we have summarized different types of physical components of vectors and tensors, scattered in earlier works in the literature \cite{truesdell1953,altman1995,frederick,mac}. We have provided an intrinsic proof
to show that all the physical components of a vector or tensor have the same physical dimension. Finally,
we use these physical components to derive the general form of the equations of elasticity which are readily
applicable for physical modelling purpose. The extension of this study to derive the hydrodynamic
equations in physical components is straightforward.

\begin{acknowledgement}
I thank the organizers of ISRA 2023. My special thanks are due to  Shubhrangshu Ghosh
and Banibrata Mukhopadhyay. The  project leading to this publication has received funding from France  2030, the French Government program managed by the French
National  Research Agency (ANR-16-CONV-0001) and from Excellence
Initiative of  Aix-Marseille University - A*MIDEX.
\end{acknowledgement}

\end{document}